\def \FigPath {./}
\documentclass[prl,twocolumn,english,superscriptaddress,floatfix]{revtex4}

\usepackage{graphicx}
\usepackage{float}
\usepackage{bm, amsmath, amssymb}
\usepackage{color}
\usepackage{soul}
\usepackage{pdfpages}
\usepackage[T1]{fontenc}

\usepackage[utf8]{inputenc}
\usepackage{braket}
\usepackage{xfrac}
\usepackage{verbatim}
\usepackage{lipsum}
\usepackage{cancel}

\usepackage{amstext}
\usepackage{bbold}
\usepackage{esint}
\usepackage{babel}
\usepackage{relsize}
\newcommand{\addkm}[1]{\textcolor{black}{#1}}
\newcommand{\addpmh}[1]{\textcolor{black}{#1}}

\begin{document}

\title{Quantum Zeno effects from measurement controlled qubit--bath interactions}

\author{P. M. Harrington}
\affiliation{Department of Physics, Washington University, St.~Louis, Missouri 63130}
\author{J. T. Monroe}
\affiliation{Department of Physics, Washington University, St.~Louis, Missouri 63130}
\author{K. W. Murch}
\affiliation{Department of Physics, Washington University, St.~Louis, Missouri 63130}
\affiliation{Institute for Materials Science and Engineering, St.~Louis, Missouri 63130}

\date{\today}

\begin{abstract}
The Zeno and anti-Zeno effects are features of measurement-driven quantum evolution where frequent measurement inhibits or accelerates the decay of a quantum state. Either type of evolution can emerge depending on the system-environment interaction and measurement method. In this experiment, we use a superconducting qubit to map out both types of Zeno effect in the presence of structured noise baths and variable measurement rates. We observe both the suppression and acceleration of qubit decay as repeated  measurements are used to modulate the qubit spectrum causing the qubit to sample different portions of the bath. We compare the Zeno effects arising from dispersive energy measurements and purely-dephasing `quasi'-measurements, showing energy measurements are not necessary to accelerate or suppress the decay process.
\end{abstract}

\maketitle

A projective measurement should reset the clock of a decay process, reinitializing the system to its excited state and therefore inhibiting decay in a variety of situations ranging from nuclear physics \cite{lane83} to optical lattices \cite{pati15}. The suppression of a decay process---and more broadly quantum evolution---by frequent measurement is referred to as the ``Zeno effect'' \cite{misr77}. The fact that the Zeno effect in decay processes is almost universally negligible is evident by considering Fermi's golden rule for determining a decay rate: the decay rate depends  on the density of states only at the transition frequency and repeated measurement effectively samples a larger range of frequencies in the calculation. \addpmh{If the bath is white on the probed band, then the decay rate is unchanged. Therefore, the Zeno effect will only occur under the special circumstance where the noise spectral density varies quickly over the probed band. Moreover}, the opposite ``anti-Zeno effect,'' where frequent measurements accelerate decay, is predicted to be a more ubiquitous phenomenon \cite{kaul97, kofm00, facc01}. Here we perform a detailed study of both Zeno effects using a superconducting qubit as an emitter coupled to a transmission line with a tunable structured bath. Frequent measurements alter the qubit--bath interaction leading to both accelerated and suppressed decay.  Our study expands on the role of measurement in the Zeno effects and highlights new ways to control quantum evolution with tunable bath interactions \cite{poya96}.

The original development of the Zeno effect predicted the inhibition of particle decay and non-exponential dynamics due to time-evolution interruption from frequent observations \cite{misr77}. The general case for any quantum system under continuous measurement, dubbed the `watchdog-effect' \cite{krau81}, was explained in terms of cancelation of wavefunction coherence caused from measurement induced perturbations, thus slowing evolution from an initial state \cite{pere80}. The first experimental measurements of the Zeno effect, conducted with trapped ions \cite{itan90}, incited much discourse on the nature of measurement, the essential features of the Zeno effect, and how the effect compares to simple perturbation dynamics due to external coupling \cite{ball91,itan91}. In recent years, the effect has been generalized as any disruption of the unitary evolution due to projection-like interactions with an external system \cite{itan06}, and it has also been suggested that Zeno-like dynamics can arise from unitary (non-projective) dynamics alone \cite{pasc94,viol98,nana01}. The anti-Zeno effect occurs when frequent measurements accelerate a decay process and was first observed (along with the Zeno effect) in a tunneling experiment with a cold atomic gas \cite{fisc01}. 

In contrast to previous work \cite{pati15,itan90,fisc01}, our experiment focuses on a single quantum system, where ensemble averaging occurs only after data collection \cite{tosc13}. While Zeno effects, and more broadly Zeno dynamics, have been studied with superconducting qubits \cite{kaku15,slic16,bret15}, the anti-Zeno effect has not yet been studied at the level of a single quantum system. In our experiment, we demonstrate how both Zeno effects arise from frequent projective energy measurements on a superconducting qubit. Furthermore, to examine the role of information for Zeno decay dynamics, we introduce a dephasing-only ``measurement'' method which does not cause measurement backaction in the energy basis.

\begin{figure}[h]
\centering
\includegraphics[angle=0,width=0.5\textwidth]{\FigPath 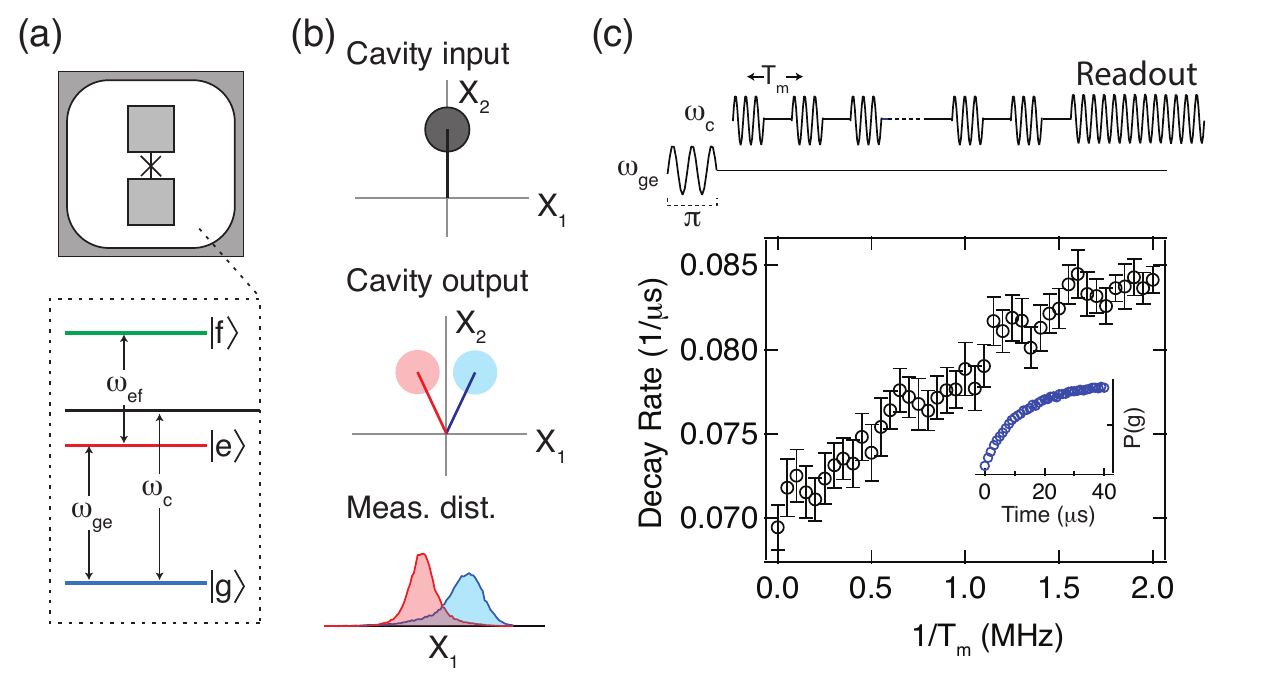}
\caption{{\bf Experimental Setup.} (a) The system consists of a transmon circuit dispersively coupled to a waveguide cavity, where the $\ket{g}$ and $\ket{e}$ states define the qubit  and $\ket{f}$ is an auxiliary state. (b) The qubit-cavity interaction results in a state-dependent phase shift on a probe near the cavity frequency, $\omega_c$, allowing the single shot measurement of the qubit state.  (c) A standard inversion recovery ($T_1$) measurement (inset) is used to characterize the effect of repeated projective measurements.  A slight increase in the decay rate as the inter-measurement time interval $T_m$ is decreased is expected from the non-QND character of the measurement.  }
\label{fig:zm}
\end{figure}

Our system consists of a transmon circuit that is dispersively coupled to a three-dimensional waveguide cavity of frequency $\omega_c/2\pi = 6.895\,\text{GHz}$ (Fig.\ 1a) \cite{koch07,paik113D}.   The two lowest energy eigenstates $\{\ket{g},\ket{e}\}$ define a qubit with a transition frequency of $\omega_{ge}/2\pi = 5.103\,\text{GHz}$. The interaction Hamiltonian, $H_\mathrm{int}/\hbar = -\chi a^\dagger a \sigma_z$ results in a qubit-state-dependent frequency shift of the cavity (Fig.\ 1b) \cite{wall05}.  Here, $a^\dagger$ $(a)$ is the creation (annihilation) operator for the cavity resonance, $\sigma_z$ is the Pauli spin operator which has the qubit energy states as an eigenbasis, and $\chi/2\pi = -1.38$ MHz is the dispersive coupling rate.  A probe that populates the cavity with an average intracavity photon number $\bar{n}$ results in dispersive measurement of the qubit energy state characterized by a timescale $\tau = \kappa/(16\bar{n} \eta \chi^2)$, where $\kappa/2\pi = 6.81$ MHz is the cavity linewidth, and $\eta  = 0.014$ is the measurement quantum efficiency \cite{murc13traj,hatr13,webe14}.  

Since the measurement operator $\sigma_z$ commutes with the Hamiltonian,  this measurement is considered Quantum Non-Demolition (QND). However, counter-rotating terms and noise mixing  break the QND character of this measurement \cite{slic12,sank16}. To  analyze repeated measurements, we perform qubit lifetime measurements while applying 100 ns long probe pulses that occupy the cavity with $\bar{n} = 9$ intracavity photons. If the probe photons were detected with unity quantum efficiency, then this measurement would distinguish between the energy eigenstates of the circuit with nearly unit fidelity.
As such, we consider these measurements to be complete projective measurements, even though the paltry quantum efficiency of the setup inhibits our ability to record these measurements with \addkm{exceedingly} high fidelity. As shown in Fig.~\ref{fig:zm}c, the observed  $T_1$ of the qubit in the presence of these measurements does decrease, though we emphasize that this alteration of the decay rate is not a Zeno effect as it can easily be explained by considering the non-QND character of the measurement \cite{slic12,sank16}.

To study Zeno effects in this system, we introduce a structured thermal bath that alters the decay rate of the transmon circuit. We synthesize a bath of a Lorentzian amplitude spectrum from a white noise source filtered by a low pass $LC$ filter resulting in a power spectrum with a 3-dB-width of $~1$  MHz. This low frequency noise is upconverted to near the qubit transition, $\omega_\mathrm{ge}$, via single sideband modulation \addpmh{(Fig.\ 2a)}. This thermal photon bath induces stimulated emission and absorption of the qubit, modifying the decay rate.  The strength of the bath coupling to the qubit transition is characterized by $N$, the average number of thermal photons. Since the bath contributes $N$ thermal photons, the radiative decay time decreases as,
\begin{equation}\label{eq:t1nth}
T_{1} = T_{1,\mathrm{spont.}} / (2N+1) \nonumber
\end{equation}
where $T_{1,\mathrm{spont.}} = 20\ \mu$s is the radiative decay time from spontaneous emission and $N$ is the number of thermal photons coupling to the qubit transition \cite{gard86,murc13}. 

\begin{figure}[h]
\centering
\includegraphics[angle=0,width=0.45\textwidth]{\FigPath 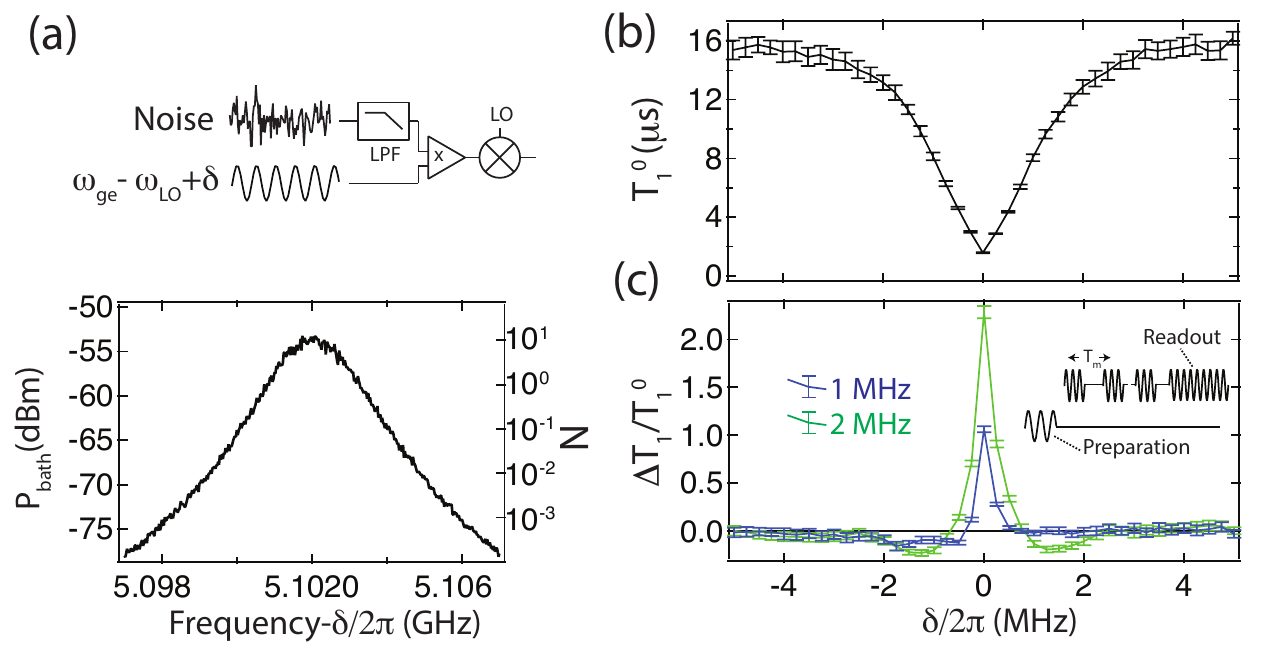}
\caption{{\bf Synthesized noise bath and Zeno effects}. (a) A structured thermal noise bath is created from a filtered white noise source mixed up to the qubit transition frequency. The bath is characterized by a Lorentzian squared power spectrum (shown here at the input plane of the dilution refrigerator), with a center frequency that is tunable from the low frequency modulation $\omega_\mathrm{ge}-\omega_\mathrm{LO}+\delta$. (b) Inversion recovery measurements in the presence of the noise bath show how the thermal photons  decrease the effective decay time when the bath is centered on the qubit transition.  \addkm{ (c) The fractional change in the qubit $T_1$ decay time, \addkm{$\Delta T_1 = (T_1-T_1^0)/T_1^0$}, versus qubit--bath detuning, where $T_1^0$ is the $T_1$ time in the absence of additional measurement, as shown in panel (b). Repeated projective measurements, applied at a rate $1/T_m = 2$ MHz (green) and $1/T_m = 1$ MHz (blue), alter the coupling of the qubit to the bath.} This either enhances (anti-Zeno effect, \addkm{below the horizontal line}) or suppresses (Zeno effect, \addkm{above the horizontal line}) the decay relative to the case without measurements.}
\label{fig:bath}
\end{figure}
To examine the effect of the synthesized noise bath on the qubit transition decay rate, we perform inversion recovery measurements for different detunings $\delta$ between the bath-center and the qubit transition. Figure \ref{fig:bath}b shows that \addkm{reducing the detuning of the center frequency} of the bath relative to the qubit transition decreases the $T_1$ coherence time according to the bath's power spectral density. We note that instead of injecting thermal photons into our system, the electromagnetic spectral density could be colored by the use of a ``Purcell filter''---a shunting narrow-band notch filter that suppresses the vacuum fluctuations \cite{reed10}.

We now focus on how repeated measurements alter the coupling of the qubit to a structured thermal environment, thereby reducing or enhancing decay. \addpmh{We can treat this with a simple theoretical model where the decay rate of the qubit is determined by the coupling of environmental modes to the qubit transition frequency. The decay rate is calculated by the overlap integral,
$$
1/T_1 = 2\pi\int_{-\infty}^{+\infty}d\omega\,F(\omega, T_m)G(\omega)
$$
where $F(\omega, T_m)$ is the qubit transition spectral profile during measurement at a rate $1/T_m$ and $G(\omega)$ is the environment density of states \cite{noripaper}. In general, the qubit spectral profile is broadened upon measurement which results in a different weighted average over the environmental modes. Consequently, when $G(\omega)$ varies in frequency, the qubit decay rate can increase or decrease due to measurement induced broadening of $F(\omega, T_m)$.} \addkm{Figure \ref{fig:bath}c displays the fractional change $\Delta T_1 = (T_1-T_1^0)/T_1^0$ of the qubit $T_1$ times versus the bath detuning frequency for different measurement rates. Here $T_1^0$ specifies the decay time for a specific bath detuning in the absence of additional measurement.} We note that the effect of an increased measurement rate is twofold: the non-QND character of higher measurement rates results in shorter $T_1$ times (as shown in Fig.\ 1), and the repeated measurements alter the coupling of the qubit to the synthesized noise bath. To isolate this second effect, we scale the measured $T_1$ values to correct for the non-QND contribution to the measured decay rate, as described in supplemental information \cite{supp}. \addkm{Since the presence of photons in the cavity causes an ac Stark shift on the qubit transition, we have also shifted the the bath detuning values slightly to make a ratiometric comparison to the no measurement case. }

As illustrated by this  comparison (Fig.~2c), the repeated measurements result in regions where the measured $T_1$ time is decreased (anti-Zeno effect, \addkm{below the horizontal line}) and increased (Zeno effect, \addkm{above the horizontal line} in Fig.\ \ref{fig:bath}c) compared to the no measurement case.


Zeno effects occur because measurement backaction perturbs the system thereby altering its coupling to the bath. The backaction of dispersive $\sigma_z$ measurements can be understood by considering the interaction Hamiltonian, $H_\mathrm{int}/\hbar = -\chi a^\dagger a  \sigma_z$. On one hand, this interaction describes an ac Stark shift of the qubit transition frequency by intracavity photons. Thus, fluctuations of the intracavity photon number leads to fluctuations of the qubit frequency and thus dephasing. On the other hand, the cavity probe accumulates information about the qubit state in the energy eigenbasis, inducing ``spooky'' backaction associated with wavefunction collapse \cite{koro11}. \addpmh{During continuous measurement} both the mechanism of photon number fluctuations and the acquired qubit state information perturb the qubit \cite{hatr13,murc13traj,Lang14}, leading to dephasing \cite{schu05,bois09}.
 The presence of these two types of backaction (pure dephasing versus eigenstate information) draws into question the role of information.  \addkm{Is projection onto an eigenstate an essential component of the Zeno effects? }Indeed, a recent proposal \cite{noripaper} has introduced the concept of a  ``quasi-measurement''---an interaction with the environment that does not necessarily accumulate information about the state---to clarify the role of wavefunction collapse in the Zeno effects.  

\begin{figure}[h]
\centering
\includegraphics[angle=0,width=0.45\textwidth]{\FigPath 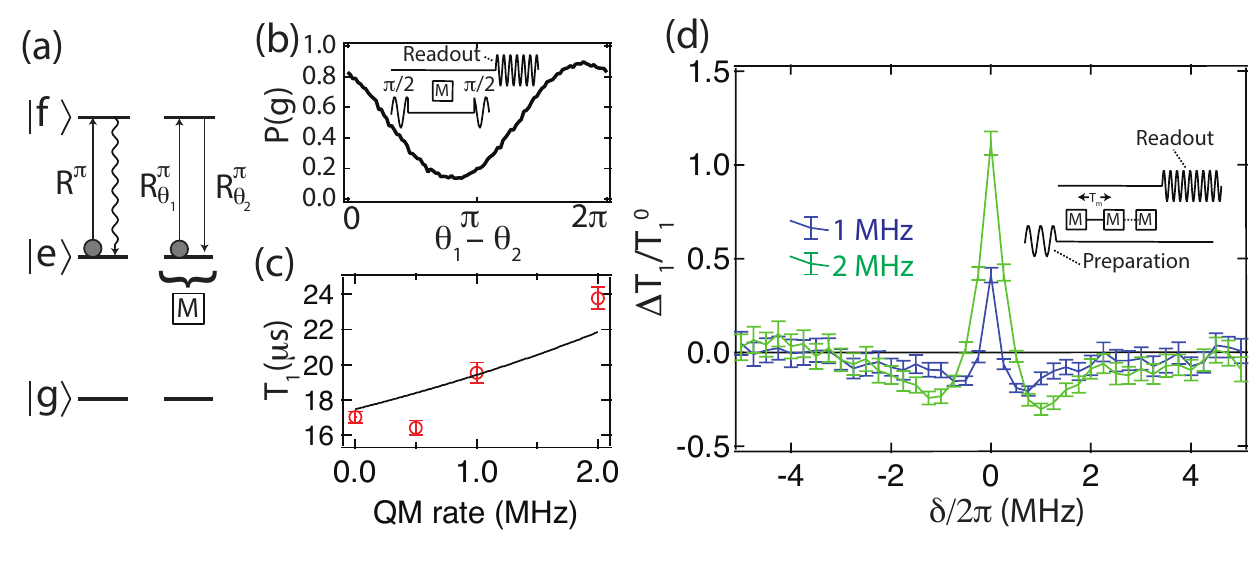}
\caption{{\bf Dephasing measurements}. (a) The proposed quasi-measurement from \cite{noripaper} involves excitation to an auxiliary state and spontaneous decay.  An alternative, dephasing measurement, uses a second $\pi$ rotation to return the system to the $\ket{e}$ state. (b) A Ramsey measurement on the qubit state to characterize the effect of the dephasing measurement. The phase difference of the two rotations in the dephasing measurement imprints Berry phase on the state $\ket{e}$. Thus, by randomizing the Berry phase, the interaction dephases the qubit. (c) Measured $T_1$ times for different measurement rates in the absence of the synthesized noise bath.  (d)  \addkm{Fractional change in the qubit} $T_1$ decay times in the presence of the synthesized noise bath for different qubit--bath detunings and different measurement rates.}
\label{fig:qm}
\end{figure}

Accordingly, we implement an alternative measurement scheme which only dephases rather than accumulates information about the qubit state. In the proposal \cite{noripaper}, a drive excites the qubit to an auxiliary state which rapidly decays through spontaneous emission (Fig.~\ref{fig:qm}a). This sequence implements a quasi-measurement: if the qubit is in the $\ket{e}$ state the system will emit a photon, making a projective measurement in the energy basis. Here, we extend this proposal to perform a dephasing-only ``measurement,'' where no information about the energy state is acquired.  To do this, we apply a  rotation $R^\pi_{\theta_1}$ on the $\omega_\mathrm{ef}$ transition to excite the circuit from $\ket{e}$ to $\ket{f}$ and then apply a second  rotation $R^\pi_{\theta_2}$ to return the circuit to the $\ket{e}$ state. These two rotations result in an accumulated Berry phase \cite{berr84,leek07,fili13,yale16} on the state $\ket{e}$. To characterize this Berry phase, we perform a Ramsey measurement as shown in Figure \ref{fig:qm}b.  The Ramsey measurements show that the rotations imprint a specific phase evolution on the qubit state related to the phase difference of the two rotations. By randomizing the Berry phase, this interaction completely dephases the qubit state. Furthermore, because the rotations are conducted with a classical drive, no information of qubit's energy state is acquired in the interaction. These ``dephasing measurements'' are similar to the dispersive energy measurements, where entanglement between the qubit and environment, and subsequent measurement of the environment, produces a random (owing to the quantum \addkm{fluctuations} of environment) perturbation on the qubit. For dephasing measurements, however, the perturbation is imprinted on the qubit by the relative phase of the rotations. The quasi-measurements do not acquire information of energy state populations, but instead only dephase the state in the energy \addpmh{eigenbasis}.

In the experiment, the dephasing measurements are implemented with two Gaussian pulses ($\sigma = 10$ ns) on the $\omega_\mathrm{ef}$ transition separated by $67$ ns. The relative phase between the two pulses is chosen from a pseudo-random number generator. Because each dephasing measurement takes the circuit out of the qubit state manifold, repeated measurements change the effective decay time as shown in Figure \ref{fig:qm}c, where the solid line indicates the expected dependence on the measured $T_1$ based on the \addkm{dephasing} measurement rate \cite{supp}.

We now return to our investigation of the Zeno effects. \addkm{ In Figure \ref{fig:qm}d, we display the fractional change in $T_1$ versus bath detuning, repeating the same experimental sequence as in Figure \ref{fig:bath}c, but we have replaced the dispersive $\sigma_z$ measurements with dephasing quasi-measurements.} Here we have our central result: the quasi-measurement scheme exhibits the same decay time pattern as dispersive energy measurements when the bath spectrum is located at specific detunings from the qubit transition. For increasing measurement rates, we find suppressed decay (Zeno-effect, \addkm{above the horizontal line}) when the bath spectrum is near the qubit transition and enhanced decay (anti-Zeno effect, \addkm{below the horizontal line}) when the bath is further detuned. At higher quasi-measurement rates the Zeno effects become more drastic. Our data show that dephasing measurements induce Zeno effects in a comparable way to projective $\sigma_z$ measurements. 

\begin{figure}[h]
\centering
\includegraphics[angle=0,width=0.45\textwidth]{\FigPath 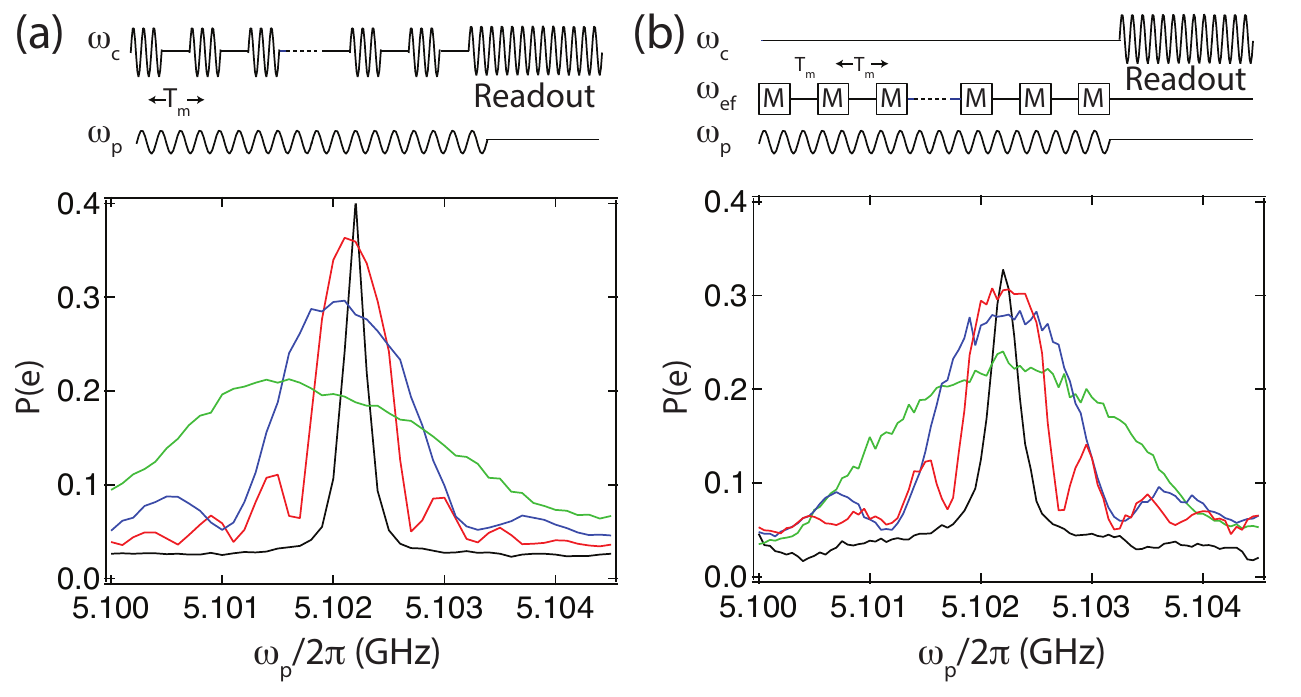}
\caption{{\bf Spectroscopy}. A continuous weak probe at frequency $\omega_\mathrm{p}$ is applied for a duration of 80 $\mu$s followed by a projective measurement. The power-broadened qubit transition frequency is revealed as an increase in the final excited state population (black traces). To probe the effects of dispersive (a) and quasi-measurements (b), we apply these measurements at different measurement rates: 0.5 MHz, 1 MHz, 2 MHz (red, blue, green, respectivley).}
\label{fig:spec}
\end{figure}

To probe how the repeated measurements alter the qubit--bath coupling we perform continuous wave spectroscopy of the qubit transition.  Accordingly, a weak probe is applied at a variable frequency for a duration of $80\,\mu\text{s}$ before a projective measurement determines the excited state population.  Figure \ref{fig:spec} shows how the final excited state population varies as a function of probe frequency for different measurement rates. By increasing measurement rates, we broaden and modulate the qubit transition. Dispersive measurements also result in a slight ac Stark shift of the transition to lower frequencies. The spectroscopy clearly shows how both dispersive and dephasing measurements perturb the qubit transition similarly, such that Zeno effects can arise depending on the spectral properties of the electromagnetic environment.

The Zeno and anti-Zeno effects occur from an emitter decoupling from or coupling to its environment. When random measurement perturbations broaden the emitter's resonance profile, the emitter samples more or less of the bath depending on the spectral density of states. Counter to the original conception of Zeno effects, the measurements that induce broadening of the emitter's transition do not need to acquire information about the system energy state, but should simply dephase the quantum state.

This experiment demonstrates tools for quantum state engineering through the interplay of radiative decay (dissipative bath interactions), dispersive  interactions (projective $\sigma_z$ measurement), and perturbation by a classical drive (dephasing from quasi-measurement). These methods can be extended to higher dimensional quantum systems to create Zeno dynamics \cite{facc02, raim10, raim12,scha13,sign14,baro15,bret15,lin16}, where measurement restricts state evolution to certain subspaces.

\begin{acknowledgements}
\emph{Acknowledgements}---We thank D. Tan for sample fabrication, as well as  M. Naghiloo, \addkm{S. Hacohen-Gourgy,} and P. Kwiat for discussions. We acknowledge primary research support NSF grant PHY-1607156 and ONR grant No. 12114811. This research used facilities at the Institute of Materials Science and Engineering at Washington University. K.W.M acknowledges support from the Sloan Foundation.
\end{acknowledgements}


\appendix

\pagebreak

\onecolumngrid
\section*{Supplementary Material}

\subsection*{Inversion Recovery Measurements and Systematic Corrections}

\begin{figure*}[h]
\centering
\includegraphics[angle=0,width=0.8\textwidth]{\FigPath 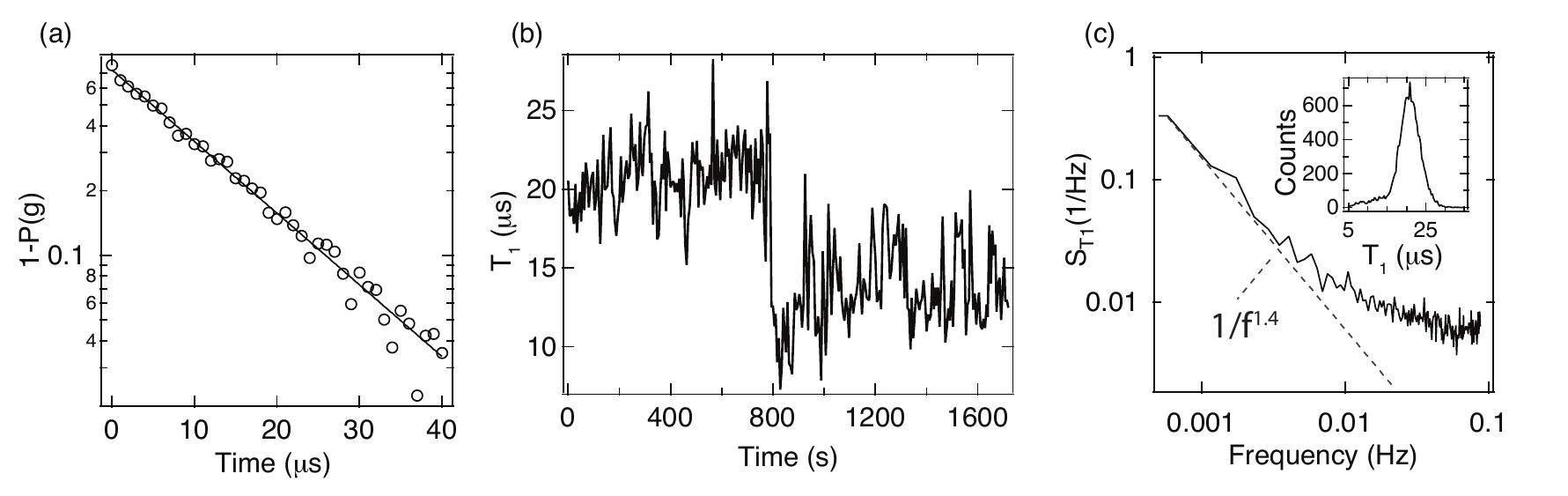}
\caption{{\bf Variation in measured $T_1$ times.} (a) Typical $T_1$ fit. (b) Variation of $T_1$ over time. (c) Power spectral density of $T_1$ fluctuations (inset: histogram of $T_1$ values obtained over a 24 hour period).}
\label{fig:t1meas}
\end{figure*}

\begin{figure}[h]
\centering
\includegraphics[angle=0,width=0.45\textwidth]{\FigPath 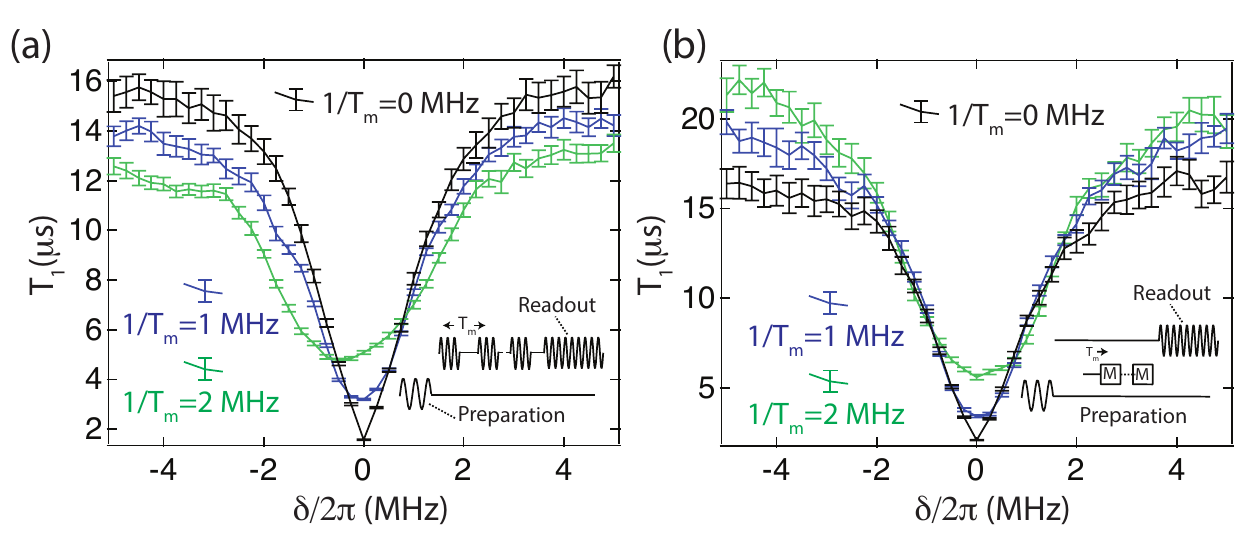}
\caption{{\bf Uncorrected $T_1$ decay curves.} (a) Higher measurement rates uniformly reduce $T_1$ for dispersive $\sigma_z$-measurements. (b) Higher measurement rates tend to increase the measured $T_1$ \addkm{with} quasi-measurements.}
\label{fig:t1uncor}
\end{figure}

Figure \ref{fig:t1meas}a displays a typical inversion recovery experiment obtained by averaging 2000 measurements at each time point. We fit the rise in ground state population rather than decay of the excited state population to avoid possible effects from coupling to the state $\ket{f}$.  Figure \ref{fig:t1meas}b displays the results of several repeated inversion recovery measurements, showing the presence of  long-timescale variations of the $T_1$ time. The power spectral density of fractional $T_1$ measurement fluctuations (Fig. \ref{fig:t1meas}c) shows that the fluctuations have a \addkm{$1/f^{\alpha}$ character with $\alpha = 1.4$}.  Figure \ref{fig:t1meas}c (inset) shows a histogram of the measured $T_1$ values indicating a mean value of $20\ \mu$s and significant variance.

The  $T_1$ versus qubit--bath detuning curves at variable measurement rates were collected in the following manner: individual inversion recovery measurements  (41 time points, 2000 measurements per point) were collected  at each of 41 different qubit--bath detunings for a given measurement rate.  These qubit--bath sweep measurements were repeated for different measurement rates. The presented data is composed of the average of 17 qubit--bath sweeps and the  error bars indicate the standard error of the mean. 

Both the dispersive energy measurements and dephasing measurements that we study in the main text introduce additional systematic effects on the  measured $T_1$ times.  In the main text, we present decay curves that are corrected for these systematic effects, and  in  Figure \ref{fig:t1uncor} we display the uncorrected $T_1$ measurements indicating that our observed Zeno effects are evident without correction.

The  non-QND character of the dispersive energy measurements is described by the  dressed-dephasing model of  \cite{slic12}, where low-frequency noise in the qubit transition is mixed with cavity photons  creating an incohoherent drive at the qubit transition.  This  effect  adds an additional decay  rate $\propto \bar{n}$, which in turn is proportional to the measurement  rate.  To correct for this effect  we use the measured $T_1$ times for large bath detuning ($|\delta|/2\pi>4$ MHz) to determine the additional decay rate arising due to the increased measurement rate and subtract this rate from the measured rates. \addpmh{These subtracted decay rates are $0.015\,\mu\text{s}^{-1}$ and $0.007\,\mu\text{s}^{-1}$ for the $2\,\text{MHz}$ and $1\,\text{MHz}$ measurement rates respectively.}  

\addkm{In addition to correcting for additional decay, we have also corrected for a slight ac Stark shift of the qubit frequency.  For the fractional change in $T_1$ presented in the main text, the $2$ MHz measurement rate data has been shifted by $+0.25 $ MHz before the fractional change is calculated.}

The dephasing measurements make use of the auxiliary state $\ket{f}$ and due to the  finite duration of the measurement operation,  $t_\mathrm{m} = 100$ ns, each measurement effectively  shortens the duration of the  inversion recovery experiment due to the  system leaving the qubit subspace. This leads to a systematic  increase in the apparent $T_1$ time, $T_{1,\mathrm{meas.}} = T_1/(1-t_\mathrm{m}/T_\mathrm{m})$.  As shown in  Figure \ref{fig:qm}c, this model roughly accounts for the observed changes in the measured $T_1$ times.  To correct for this effect  we use the measured $T_1$ times for large bath detuning ($|\delta|/2\pi>4$ MHz) to determine the linear correction  factor for each measurement rate and the $T_1$ decay curves are scaled accordingly. \addkm{These $T_1$ scaling correction factors are $0.867$ and $0.637$ for the $1\,\text{MHz}$ and $2\,\text{MHz}$ measurement rates respectively.}


\begin{figure}[h]
\centering
\includegraphics[angle=0,width=0.5\textwidth]{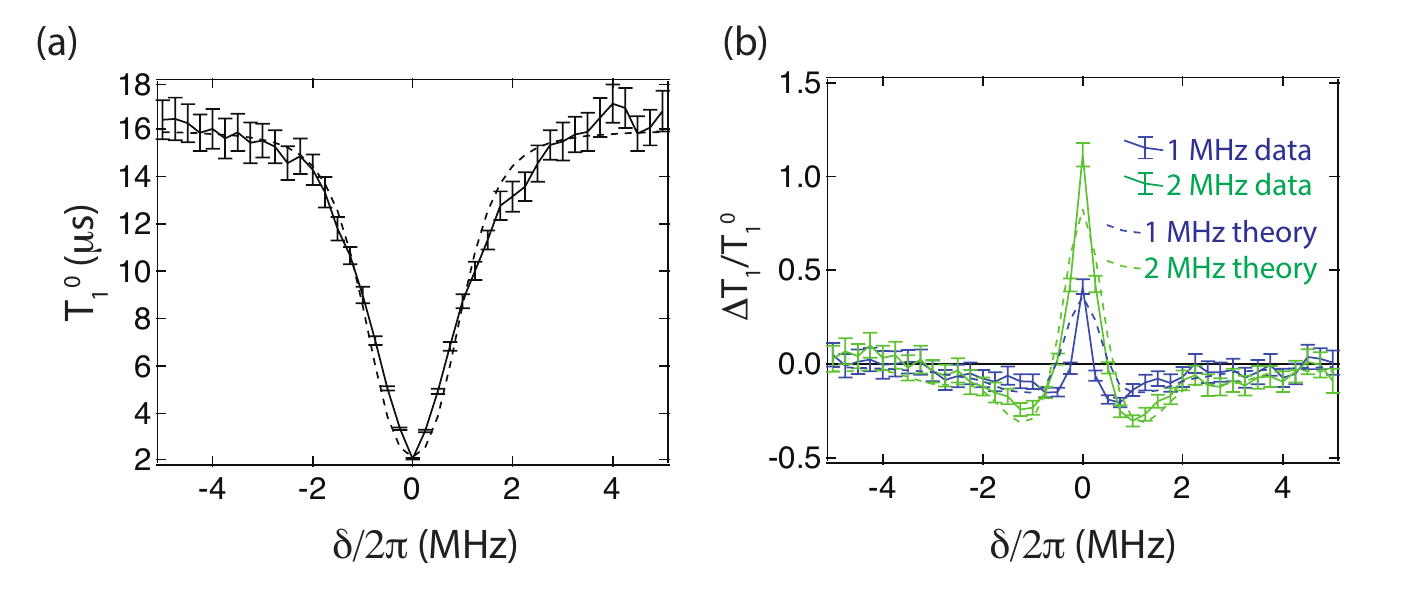}
\caption{{\bf Comparison between theory and experiment.} (a) \addkm{The bath detuning dependence of the $T_1^0$ decay time without measurement (solid line) compared to the decay rate given a noise bath (Eq.\ (\ref{eq:bath}) of width $B/2\pi = 0.78$ MHz. }  (b) Fractional change in $T_1$ for different measurement rates. The solid lines are the data presented in Figure 3 and the dashed lines are the respective theory curves based on  Eq.\ (\ref{eq:thy}).}
\label{fig:theory}
\end{figure}

\subsection*{Decay Rate Calculation}
The decay rate of an emitter results from coupling to environmental modes at the transition frequency. The decay rate is calculated by the overlap integral,
\begin{equation}
1/T_1 = 2\pi\int_{-\infty}^{+\infty}d\omega\,F(\omega, T_m)G(\omega) \label{eq:thy}
\end{equation}
where $F(\omega, T_m)$ is the qubit transition spectral profile during measurement at rate $T_m$ and $G(\omega)$ is the environmental density of states \cite{noripaper}.  In our experiment the power spectrum of the thermal noise bath is given by, 
\begin{equation}
G(\omega) = \left(\frac{A}{(\omega-\delta)^2 + B^2}\right)^2, \label{eq:bath}
\end{equation}
where  $\delta$ is the qubit-bath detuning, $A$ characterizes the strength of the bath, and $B/2\pi = 1.3$ MHz  characterizes the width.  Under the approximation that the measurements are instantaneous and projective,  qubit transition spectral profile  is given by  \cite{noripaper},
\begin{equation}
F(\omega, T_m) =T_m \left(\frac{\sin(\omega T_m/2)}{\omega T_m/2}\right)^2.
\end{equation}
\addkm{To compare this simple theoretical model to our data, we use the measured decay times versus bath detuning to determine the spectrum of the bath. From this we determine that the bath spectrum is a Lorentzian (Eq.\ (\ref{eq:bath})) of width $B/2\pi = 0.78$ MHz. This width is slightly narrower than the width that is recorded in the noise power spectrum at room temperature ($B/2\pi = 1.3$ MHz). Figure \ref{fig:theory}(a) compares the measured $T_1$ decay times in the absence of measurement to the decay calculated by Eqs. (\ref{eq:thy}) and (\ref {eq:bath}).   Figure \ref{fig:theory}(b) compares the experimental data of Figure 3 (main text) to the calculated fractional change in $T_1$ showing good agreement with theory.}

\subsection*{Fixed Phase Quasi-Measurements}
\addkm{As discussed in the introduction, Zeno-like effects can arise from unitary driving alone \cite{pasc94,viol98,nana01}. Here we show how quasi-measurements, without phase randomization, split the qubit transition due to repeated unitary rotations. The qubit line does not broaden because, without phase randomization, the qubit state retains coherence. Since Zeno effects arise from measurement altered coupling to the thermal bath, we still see Zeno-like effects even without dephasing.}
\begin{figure}[h]
\centering
\includegraphics[angle=0,width=0.75\textwidth]{\FigPath 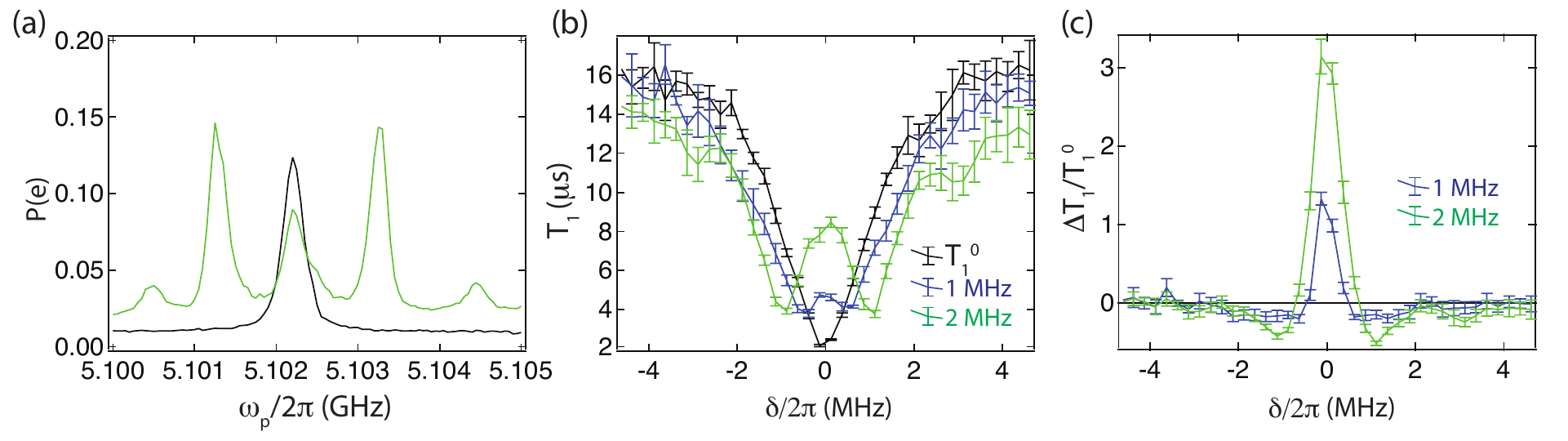}
\caption{\addpmh{{\bf Fixed Phase Quasi-Measurements.} (a) Spectroscopy for the case of no measurement (black) and the case of fixed phase quasi-measurements at a rate of 2 MHz (green). (b) The uncorrected $T_1$ decay times versus bath detuning. (c) The fractional change in $T_1$ as compared to the no measurement case.}}
\label{fig:supp5}
\end{figure}

\subsection*{Experimental Apparatus}

\begin{figure}[h]
\centering
\includegraphics[angle=0,width=0.6\textwidth]{\FigPath 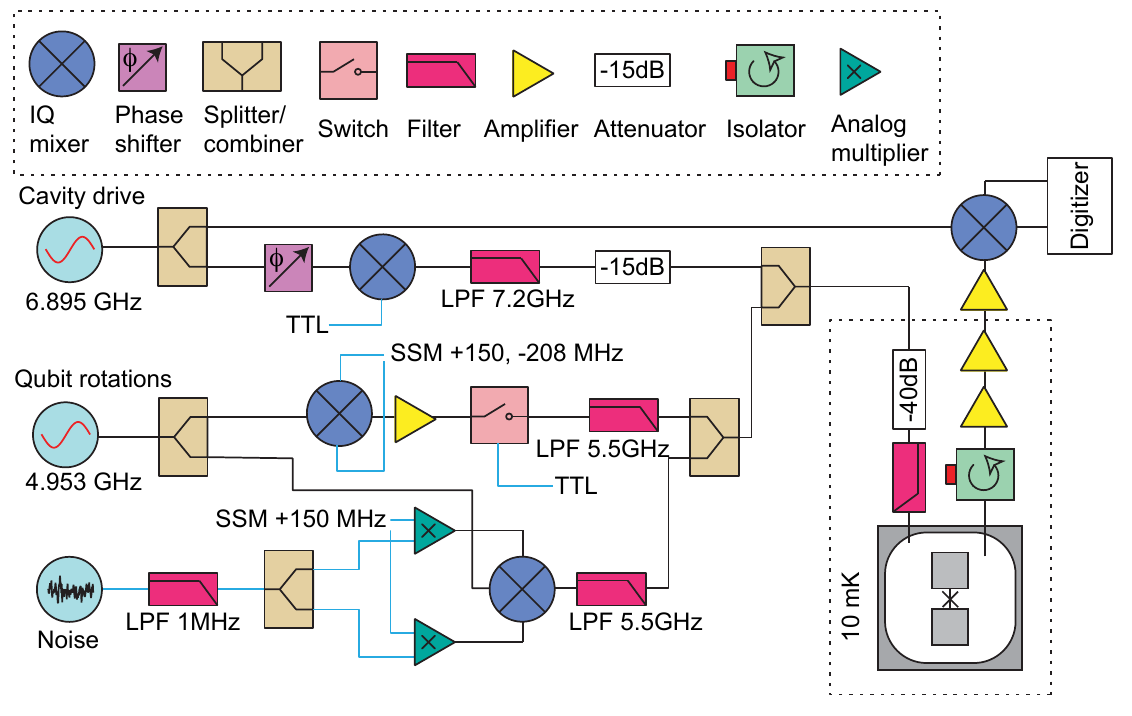}
\caption{{\bf Experimental schematic.} State manipulations on the transmon circuit are performed with single sideband modulation. The thermal noise bath is created by mixing low frequency noise up to near the qubit frequency with single sideband modulation.}
\label{fig:setup}
\end{figure}

\end{document}